\documentstyle[psfig]{mn}

\def\spose#1{\hbox to 0pt{#1\hss}}
\def\lta{\mathrel{\spose{\lower 3pt\hbox{$\mathchar"218$}}
     \raise 2.0pt\hbox{$\mathchar"13C$}}}
\def\gta{\mathrel{\spose{\lower 3pt\hbox{$\mathchar"218$}}
     \raise 2.0pt\hbox{$\mathchar"13E$}}}
\def\Msun{{\rm M}_\odot}

\def\src{4U~1708--40 }
\def\sr{4U~1708--40}

\begin{document}

\title[Discovery of type-I X-ray bursts from the LMXB 4U~1708--40]
{Discovery of type-I X-ray bursts from the low-mass X-ray binary 4U~1708--40}

\author[S.~Migliari et al.]
{S. Migliari$^1$\thanks{email: migliari@astro.uva.nl},
T. Di~Salvo$^1$, T. Belloni$^2$, M. van der Klis$^1$,
R. P. Fender$^1$, 
\newauthor
S. Campana$^2$, C. Kouveliotou$^{3,4}$, M. M\'endez$^5$, W. H. G. Lewin$^6$ \\
$^1$ Astronomical Institute ``Anton Pannekoek'' University of Amsterdam and
Center for High-Energy Astrophysics, Kruislaan 403,\\
1098 SJ, Amsterdam, The Netherlands\\ 
$^2$ INAF--Osservatorio Astronomico di Brera, Via E. Bianchi 46, I-23807
Merate (LC), Italy\\
 $^3$ NASA Marshall Space Flight Center, SD50, Huntsville, AL 35812.\\
$^4$ Universities Space Research Association/NSSTC, SD-50, Huntsville, AL
35805\\ 
$^5$ SRON, National Institute for Space Research, Sorbonnelaan 2, 3584 CA,
Utrecht, The Netherlands\\
$^6$ Center for Space Research and Department of Physics, Massachusetts
Institute of Technology,70 Vassar Street, \\
Cambridge, MA 02139-4307\\
}

\maketitle

\begin{abstract}

We report the discovery of type-I X-ray bursts from the
low-mass X-ray binary 4U~1708--40 during the 100~ks observation
performed by BeppoSAX on 1999 August 15--16.
Six X-ray bursts have been observed.
The unabsorbed 2--10~keV fluxes of the bursts range from
$\sim$ ($3-9$) $\times 10^{-10}$ erg~cm$^{-2}$s$^{-1}$.   
A correlation between peak flux and fluence of the bursts is found, in
agreement with the behaviour observed in other similar sources. 
There is a trend of the burst flux to increase with the time interval from the
previous burst.
From the value of the persistent flux we infer a mass
accretion rate ${\dot M}\sim 7\times 10^{-11}~\Msun$/yr, that
may correspond to the mixed hydrogen/helium burning regime triggered by
thermally unstable hydrogen. 
We have also analysed a BeppoSAX observation performed on 2001 August 22 and
previous RXTE observations of 4U~1708--40, where no
bursts have been observed; we found persistent fluxes of more than a
factor of $7$ higher than the persistent flux observed during the BeppoSAX
observation showing X-ray bursts. 

\end{abstract}


\section{Introduction}

Many low-mass X-ray binaries (hereafter LMXBs) are known to show type-I
X-ray bursts 
which have proven to be important diagnostic tools for these systems (see
Lewin, van Paradijs \& Taam 1995 for a review).  
X-ray bursts are thought to originate from thermonuclear flashes, caused by
thermal instabilities, in the accreted matter on the surface of weakly
magnetized neutron stars (NS). 
The time interval between the bursts ranges from
tens of minutes to days. In a source the burst activity can stop 
for periods from days to months (but also years, as, for example, in the case
of transient sources). In some cases, a relation has
been observed between burst properties and persistent flux. 
As the persistent flux increases the 
recurrence time between bursts tends to increase (typically from
a few hours to more than a day) and the bursts become less energetic
(e.g., Hoffman, Lewin \& Doty 1977).
For instance, in the case of 4U~1705--44 the persistent flux increased by 
a factor $\sim 2$ while the burst intervals increased by a factor of $\sim 4$
(Langmeier et al. 1987; Gottwald et al. 1989). When the persistent flux
increases further the bursts disappear completely.
 This behaviour is observed in several sources like, for example,
MXB~1659--29  (Lewin, Hoffman \& Doty 1976), GX~3+1 (Makishima et al. 1983)
and EXO 0748--673 (Gottwald et al. 1986). 
However, in other sources
(e.g. 4U~1820--30: Clark et al. 1977; 4U~1728--34: Basinska et al. 1984) the 
opposite behaviour was sometimes observed: the recurrence time between  
bursts decreased while the persistent flux increased. Nevertheless, at least
in the case of 4U 1820--30 (Clark et al. 1977), the bursts (again) disappear
for higher acretion rates, when the source is in its high state.      

Current theories of type-I X-ray bursts (Fujimoto et al. 1981; Fushiki and
Lamb 1987; see Bildsten 2000) predict three different regimes
in mass accretion rate for unstable burning:  
i) mixed H/He burning triggered by thermally unstable H ignition,  
at low accretion rates ($\dot M < 2
\times 10^{-10}$ $M_\odot$/yr); ii) pure He shell ignition following steady H 
burning, at intermediate accretion rates ($2 \times 10^{-10}$ 
$M_\odot$/yr $< \dot M <$ $4-11 \times 10^{-10}$ $M_\odot$/yr); 
iii) mixed H/He burning triggered by thermally unstable He ignition at
high accretion rates ($4-11 \times 10^{-10}$ $M_\odot$/yr $< \dot M <$ 
$2 \times 10^{-8}$ $M_\odot$/yr). At $\dot M$ even higher steady He
burning occurs, the nuclear fuel for the burst is depleted, causing the 
bursts to disappear completely.        
On the basis of these theories, an anticorrelation 
between persistent flux and burst recurrence time is
expected; since as $\dot M$
increases a lower amount of time is needed to 
accumulate the critical amount of fuel necessary to begin the burst,
the recurrence time between bursts is expected to decrease with increasing
$\dot M$. This is contrary to what is observed (e.g., Hoffman, Lewin \& Doty
1977; van~Paradijs, Penninx \& Lewin 1988).
There are basically two interpretations for these observations.  
van Paradijs et al. (1988) found evidence of a
continuous stable burning of accreted nuclear fuel in burst sources; the
observed correlation between persistent emission and burst recurrence time can
be explained as the rate of this stable burning, increasing, removes nuclear
fuel for thermonuclear flash.
Recently, Bildsten (2000) proposed that another solution of this mismatch
between theory and observations can be obtained by relaxing the hypothesis
of spherical accretion and assuming that the accretion of fresh material
only occurs on a fraction (e.g., the equator) of the NS surface. The crucial
parameters, in this case, is the accretion rate per unit area: if the covered
area increases enough with increasing $\dot M$, the accretion rate per
unit area indeed decreases.
Indeed, the whole scenario is not completely clear yet, since recent
results (e.g. Muno et al. 2000; Franco 2001; van Straaten et al. 2001) suggest
that there is not a unique trend between $\dot M$ and burst properties for all
the sources.
    
Type-I X-ray burst profiles depend strongly on photon energy: decays are much 
shorter at high photon energies than at low photon energies (Lewin et
al. 1995). This softening of the burst spectrum during the decay 
results from the cooling of the NS photosphere. 
If the luminosity during the burst reaches the Eddington limit
$L_{Edd}$, the NS photosphere expands. Since for a blackbody (see below) $L_B
\propto R^2 T_{eff}^4$, when the radius of the photosphere R expands,
the effective temperature $T_{eff}$ decreases (Tawara et al. 1984; Lewin,
Vacca \& Basinska 1984; Vacca, Lewin \& van Paradijs 1986).  

In this paper we report on the discovery of type-I X-ray bursts from the
LMXB 4U 1708--40 (Forman et al. 1978) during a BeppoSAX
observation. Not much is known about this system. A first analysis of the
2--12~keV  X-ray spectrum of 4U 1708--40 was made by Warwick et al. (1988)
using EXOSAT observations during which the 2--6~keV source flux was $\sim
10^{-9}$~erg s$^{-1}$ cm$^{-2}$. The spectrum was well fitted by a power law
with a spectral index $\Gamma=2.2\pm0.2$ and a column density N$_{\rm H}\sim
3\times 10^{22}$~cm$^{-2}$.    

\begin{table}
\centering
\caption{Best-fit parameters and 1$\sigma$ errors of the spectra of the
persistent emission during the 1999 August 15--16 (SAX 1999), 2001 August 22
(SAX 2001) BeppoSAX  observations (0.12--10~keV) and the 2000 June 18 PCA/RXTE
observation (2.5--25~keV). The 1999 BeppoSAX data are fitted with an
absorbed power law or a CompTT model plus a Gaussian emission line, the 2001
BeppoSAX data with an absorbed power law plus a blackbody component and the
RXTE data with an absorbed cutoff power law plus a blackbody. N$_{{\rm H}}$ is
the equivalent hydrogen column density in units of $10^{22}$ cm$^{-2}$,
N$_{{\rm pl}}$ is the normalisation of the power law in units of
$10^{-2}$ photons/keV/cm$^{2}$/s at 1~keV, $\Gamma$ is the photon index of the
power law, T$_{0}$ is the temperature of 
the soft seed photons for the Comptonisation, T$_{e}$ is the electron
temperature, $\tau$ is the optical depth of the spherical scattering cloud,
N$_{{\rm TT}}$ is the CompTT normalisation ($\times 10^{-2}$) in XSPEC v11.1.0
units (see Titarchuk 1994a,b), N$_{{\rm cut}}$ is the normalisation of the
cutoff power law in units of $10^{-1}$ photons/keV/cm$^{2}$/s at 1~keV,
E$_{\rm cut}$ is the cutoff energy, N$_{{\rm bb}}$ is the normalisation of the
blackbody in
units of L$_{37}$/D$^{2}_{10}$ (where L$_{37}$ is the 
source luminosity in units of $10^{37}$~erg/s and D$_{10}$ is the distance of
the source in units of 10~kpc) and T$_{\rm bb}$ the color temperature of the
blackbody. E$_{line}$ is the centroid energy,
$\sigma_{line}$ is the width and eqw is the equivalent width of the Gaussian
emission line. F$_{tot}$ is the unabsorbed 2--10~keV flux in erg s$^{-1}$
cm$^{-2}$.}  
\label{tab_pers}
\vspace{0.2cm}
\begin{tabular}{l|l l l}
\hline
\hline
SAX & Parameters  & Power law  & CompTT     \\
1999&             & + Gaussian & + Gaussian \\
\hline
& N$_{{\rm H}}$              & $2.93\pm0.08$  & $2.45\pm0.14$\\
& $\Gamma$                   & $2.42\pm0.02$  & $\ldots$\\
& N$_{{\rm pl}}$ (LECS)      & $8.54\pm0.34$  & $\ldots$\\
& N$_{{\rm pl}}$ (MECS)      & $10.72\pm0.31$ & $\ldots$\\
& kT$_{0}$ (keV)             & $\ldots$       & $0.32\pm0.03$\\
& kT$_{e}$ (keV)             & $\ldots$       & $5.30\pm0.9$\\
& $\tau$                     & $\ldots$       & $3.3\pm 0.3$\\
& N$_{{\rm TT}}$ (LECS)      & $\ldots$       & $1.4\pm0.2$\\
& N$_{{\rm TT}}$ (MECS)      & $\ldots$       & $1.7\pm0.2$\\
& E$_{line}$ (keV)           & $6.5\pm0.1$    & $6.5\pm0.1$\\   
& $\sigma_{line}$ (keV)      & $0.9$          & $0.9$ \\  
& eqw  (eV)                  & 344            & 286   \\  
& F$_{tot}$                  & $1.2\times10^{-10}$  & $1.2\times10^{-10}$  \\
\hline
\hline
SAX & Parameters  & Power law  &      \\
2001&             & + blackbody &      \\
\hline
& N$_{{\rm H}}$              & $2.9$ (fixed) & \\
& $\Gamma$                   & $2.68\pm0.29$ & \\
&  N$_{{\rm pl}}$ (MECS)     & $41.88\pm9.09$ & \\
& kT$_{\rm bb}$ (keV)        & $1.31\pm0.03$ &   \\
& N$_{{\rm bb}}$ (MECS)      & $1.02\pm0.11$ &   \\
& F$_{tot}$                  & $8.9\times10^{-10}$  &   \\         
\hline
\hline
RXTE & Parameters & Cutoffpl    & \\
     &            & + blackbody & \\
\hline
& N$_{{\rm H}}$              & $2.9$ (fixed) & \\
& $\Gamma$                   & $1.44^{+0.13}_{-0.22}$  &   \\
& N$_{{\rm cut}}$            & $3.30\pm0.38$ &   \\
& E$_{\rm cut}$ (keV)        & $4.84\pm 0.36$ &   \\ 
& kT$_{\rm bb}$ (keV)        & $1.27\pm0.02$ &   \\
& N$_{{\rm bb}}$             & $0.36\pm0.03$ &   \\
& F$_{tot}$                  & $7.4\times10^{-10}$     &   \\
\end{tabular}
\end{table}

\section{Observations and Data Analysis}

\begin{figure}
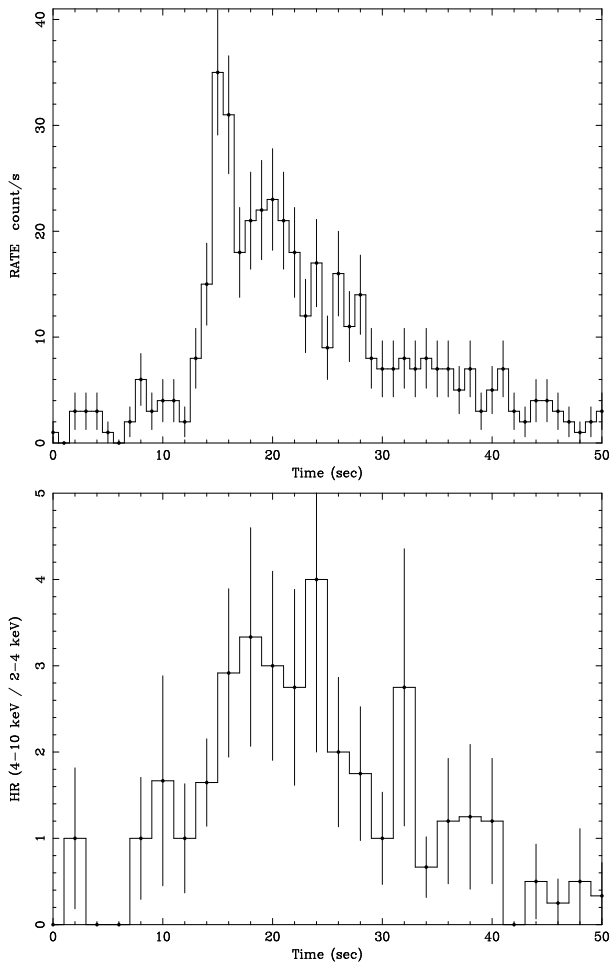

\centering
\begin{tabular}{c}
\psfig{figure=./mecs_licu_burst1.ps,width=8cm,angle=-90,height=6.3cm}\\
\psfig{figure=./mecs_HR_burst1_err.ps,width=8cm,angle=-90,height=6.3cm}\\
\end{tabular}
  \caption{{\it Upper panel}: light curve with 1~s time resolution of the
first burst. {\it Lower panel}: hardness ratio (HR$=4-10$
keV$/2-4$ keV) curve of the first burst with 2~s resolution. The shape
of the burst, the hardening during the rise and the softening during the decay
are typical of a type-I X-ray burst.}
  \label{licuB1}
\end{figure}

We have analysed a BeppoSAX observation of 4U 1708--40, performed on
1999 August 15--16 for a total observation time of $\sim 100$~ks. 
We report on results from the Low Energy Concentrator Spectrometer
(LECS, a thin-window position-sensitive gas scintillator proportional counter
with extended low energy response, 0.1--10~keV, and a field of view, FOV, of
$20^{\prime}$; Parmar et al. 1997) and the Medium Energy
Concentrator Spectrometers (MECS, position sensitive
gas scintillator proportional counters operating in the 1.3--10.5~keV
band, with a FOV of $30^{\prime}$; Boella et al. 1997b).
The source was not detected in the Phoswich Detection System (PDS, energy
range 13--200 keV; Frontera et al. 1997).
For both imaging instruments we selected the data in circular
regions of radius $8^{\prime}$ centered on the source.   
We used the standard response matrices and background files (1997 version for
MECS and 2000 for LECS) for the spectral analysis.   
To analyse the BeppoSAX spectra of the source we selected the energy range
0.12--4 keV for LECS and 1.8--10 keV for MECS. 
In the 1999 August 15-16 BeppoSAX observation six X-ray
bursts are detected. We have used LECS and MECS combined to analyse the
persistent emission, the second, third, fourth and fifth bursts, and only the
MECS to analyse the first and the sixth bursts, since the LECS was off in
these time intervals. The BeppoSAX observation is periodically
interrupted due to earth occultations and the passage through the South 
Atlantic Anomaly. These `gaps' last on average $\sim2$~ks after 
$\sim4$~ks of on-source observation. The BeppoSAX instruments 
were off more often in the last part of the observation  
(namely between the fourth and the sixth burst) so that the on-source exposure
time is $\sim 50$~ks.                

We have also analysed a 150~ks BeppoSAX observation of 4U 1708--40 performed on
2001 August 22 (with an on-source exposure time of $\sim 73$~ks) and 14
(non contiguous) observations from the public RXTE data archive, five
performed in 1997    
and nine in 2000, with a total on-source exposure time of $\sim 60$~ks.   
For the RXTE observations we have used data taken with the Proportional
Counter Array (PCA, which consists of five co-aligned Proportional Counter
Units, PCUs, sensitive 
in the energy range 2--60 keV; Zhang et al. 1993). Starting from 2000 May 12,
the propane layer on Proportional Counter Unit 0 (PCU0), which functions as an
anticoincidence shield for charged particles, was lost. Therefore we excluded
the PCU0 data from the 2000 June spectrum we have analysed. 
The PCA observations background, estimated using {\tt pcabckest} v2.1e was
also subtracted.  

We produced light curves of the 2001 BeppoSAX
observation and of all the RXTE observations; in all these
data we did not find any burst. 
Although a small variation in average count rate (around 20\%) occurred
in the RXTE data between the 1997 observations and the 2000 observations, the
light curve of each observation shows steady persistent emission without any
significant variation in count rate; we analysed the spectrum of just one of
the RXTE observations available, as a representative case.  
To analyse the PCA spectrum we have used {\tt Standard2} data in the energy
range 2.5--25 keV and produced the detector response matrix with
{\tt pcarsp} v7.11. A systematic error of 1\% was
added to account for residual uncertainties in the detector calibration.
We have used XSPEC v11.1.0 to fit the spectra.

\section{Results}

\begin{figure}
\centering
\begin{tabular}{c}
\psfig{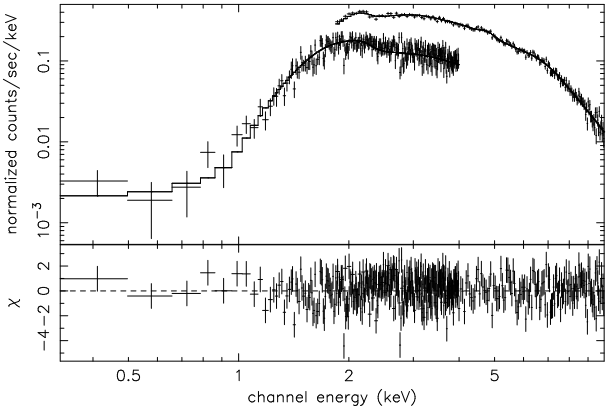}
\end{tabular} 
  \caption{{\it Upper panel}: spectrum from BeppoSAX data (LECS and MECS) of
the persistent emission of 4U~1708--408 in the range 0.12--10 keV, averaged
over more than 45~ks of observation, and the best-fit model (absorbed power
law) reported in Table~1. {\it Lower panel}: residuals in units of $\sigma$
with respect this model.}  
  \label{pers_spec_sax}
\end{figure}

\begin{figure}
\centering
\begin{tabular}{c}
\psfig{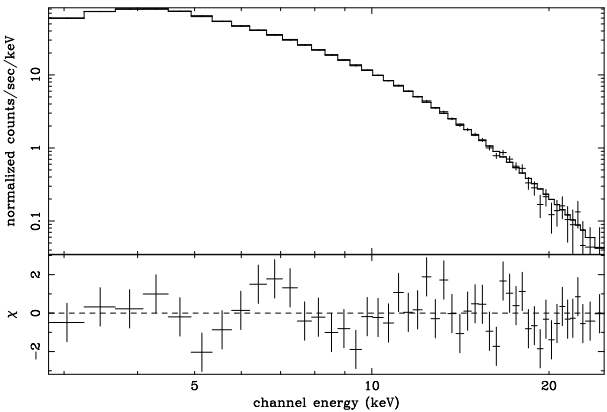}
\end{tabular} 
  \caption{{\it Upper panel}: RXTE/PCA spectrum of the persistent emission
of 4U~1708--408 in the 2.5--25 keV energy band, with the best-fit model
(absorbed cutoff power law and blackbody, Table~1; residuals around
6--7~keV are still present: see \S~3.1). The spectrum is averaged    
over the whole $\sim 3$~ks observation of 2000 June 18. {\it Lower panel}:
residuals in units of $\sigma$ with respect this model. }
  \label{pers_spec_rxte}
\end{figure}

In the 1~s time resolution LECS and MECS light curves of the 1999
observation we see persistent emission, at a count rate of $\sim 2-4$ counts/s
in the MECS light curve. 
The light curve also shows six X-ray bursts (the times at
which each burst occurred are reported in Table~\ref{tabspecB}).  
The bursts last $\sim 40$~s. Excluding the bursts, the persistent
emission does not show significant intensity variations during the whole 
observation. Five bursts out of six exhibit a rapid rise ($\la 5$
sec) followed by a slower decay, whereas one of the bursts (the
fourth) has a triangular shape with approximatly equal rise and decay times of
20 sec. In Fig.~\ref{licuB1} we show the light curve 
in the MECS range (upper panel) and the corresponding hardness
ratio (HR$=4-10$ keV$/2-4$ keV, bottom panel) of the first burst, the one
observed with the best statistics (the behaviour of the other bursts are
comparable with this). The hardening
during the rise and the softening during the decay, together with the shape
of the bursts, identify \src as a type-I X-ray burster.

\subsection{Persistent Emission}\label{persistent}

We produced LECS and MECS spectra of the persistent emission of the 1999
August 15--16 observation of 4U 1708--40 excluding intervals of 200~s
around (starting about 70~s before) each of the six detected bursts. The
spectrum in the range 0.12--10~keV is shown in Fig.~\ref{pers_spec_sax}. 
We fit the spectrum with a power law corrected for photoelectric absorption
and a Gaussian emission line at $6.5\pm0.1$ keV
which gives a reduced $\chi^{2}$ of $1.2$ for 420 $d.o.f.$  (the reduced
$\chi^{2}$ without the Gaussian emission line is 1.3 for 423 $d.o.f.$). We
find a high equivalent hydrogen column density, N$_{\rm H}= (2.93\pm
0.08)\times10^{22}$ cm$^{-2}$, consistent with the position of the source in
the direction of the galactic center, and a power law photon 
index $\Gamma=2.42\pm 0.02$. No significant thermal component ({\it i.e.},
blackbody) is found (with an upper limit on the flux of $\sim 2\times
10^{-12}$ erg~cm$^{-2}$s$^{-1}$, fixing the blackbody temperature to 1.3~keV,
see below). The 2--10 keV flux is $1.2\times10^{-10}$
erg~cm$^{-2}$s$^{-1}$, corresponding to a luminosity of $\sim 9\times10^{35}$
erg~s$^{-1}$ at a distance of 8~kpc (see Section~4).  
We obtain a fit of similar quality (reduced $\chi^{2}=1.2$ with 421 {\it
d.o.f.}) using the Comptonisation model {\it CompTT} (Titarchuk 1994a,b),
with a slightly lower column density, 
N$_{\rm H}= (2.56\pm 0.14)\times10^{22}$ cm$^{-2}$, and a Gaussian emission
line at $6.5\pm0.1$ keV. In Table~\ref{tab_pers} we
show the fit parameters of the persistent emission corresponding to these two
models.

For comparison we have also analysed the \src spectrum during one of the RXTE
observations (the 3~ks observation performed in 2000 June 18, obsID
50060-04-01-00). A simple absorbed (the column density
was fixed to N$_{\rm H}=2.9\times10^{22}$ cm$^{-2}$, in accordance with the
BeppoSAX spectrum) power law is not able to fit the spectrum in the whole
energy range. A  blackbody component with a
color temperature of $\sim 1.3$~keV 
is needed to adequatly fit the data (an F-test gives a probability 
of chance improvement of the fit for the addition of this component of
$2\times10^{-7}$). A high energy cutoff at relatively low
energy (E$_{cut}\sim 5$~keV) is also necessary (reduced $\chi^{2}=0.9$
with 44 {\it d.o.f.}). We show this spectrum,
together with residuals in units of $\sigma$ with respect to the best-fit
model, in Fig.~\ref{pers_spec_rxte} and the best-fit parameters in
Table~\ref{tab_pers}. 
We measure a 2--10 keV unabsorbed flux of $\sim 7.4\times10^{-10}$~erg
cm$^{-2}$s$^{-1}$, a factor $\sim 7$ times higher than the flux of the
persistent emission we measure during the BeppoSAX observation.
In Fig.~\ref{pers_spec_rxte} we note residuals around 6--7~keV. For
comparison with the spectrum during the 1999 August 15--16 BeppoSAX 
observation, we tried to add a Gaussian emission line in the
range 6.4--6.9 keV. We find a Gaussian line at $\sim 6.75$ keV
with an equivalent width of 87 eV.    
The parameters of the other components does not change significantly.  

We have also analysed the MECS spectrum of the BeppoSAX observation performed
on 2001 August 22. The best-fit is obtained using a
power law with a slope of $\Gamma=2.68\pm0.28$, and a blackbody with a
color temperature of kT$=1.31\pm0.03$ keV (reduced $\chi^{2}=1.07$ with 172
d.o.f.; Table~\ref{tab_pers}).
The spectrum is compatible with a Gaussian emission line between 6.4 and
6.7 keV, similar to the one used to fit the source spectrum during the
previous BeppoSAX and RXTE observations, although this component is not
required in this case. The unabsorbed 2--10~keV
flux is $8.9\times10^{-10}$~erg cm$^{-2}$s$^{-1}$. This  model is consistent
with the best-fit model of the PCA 2--10~keV spectrum, that is well fit
with a power law with a $\Gamma\sim2.77$ and a blackbody with a color
temperature of $\sim 1.3$~keV.

\begin{table}
\centering
\caption{Blackbody color temperature (kT$_{{\rm bb}}$ in keV), normalisation
(N$_{bb}$, in units of L$_{37}$/D$^{2}_{10}$ where L$_{37}$ is the 
source luminosity in units of $10^{37}$~erg/s and D$_{10}$ is the distance to
the source in units of 10~kpc)
and 2--10~keV unabsorbed flux F ($\times 10^{-10}$ erg cm$^{-2}$s$^{-1}$)
with $1\sigma$ errors,  
from the fit of the averaged spectra over 40~s of the six bursts with a
blackbody component.  
In the first column in brackets is also indicated the time (with 1~s time
resolution) from the beginning of the MECS observation (15 August 1999 at
05:01:10.5) at which the peak of each burst occurs.}
\label{tabspecB}
\vspace{0.2cm}
\begin{tabular}{l|l l l }
\hline
\hline
 burst \# (time)& kT$_{{\rm bb}}$  & N$_{{\rm bb}}$   & F    \\
\hline				                                          
1 (21655 s) & $2.01\pm0.16$ & $1.52\pm0.17$ & $9.2\pm0.6$ \\
2 (29160 s) & $1.60\pm0.13$ & $0.75\pm0.08$ & $5.6\pm0.4$ \\
3 (39620 s) & $1.64\pm0.10$ & $1.12\pm0.09$ & $7.7\pm0.5$ \\
4 (52350 s) & $1.75\pm0.12$ & $1.49\pm0.14$ & $9.6\pm0.7$ \\
5 (57498 s) & $1.28\pm0.09$ & $0.68\pm0.06$ & $5.1\pm0.4$ \\
6 (91064 s) & $1.40\pm0.15$ & $0.45\pm0.06$ & $3.8\pm0.3$ \\

\end{tabular}
\end{table}

\subsection{X-ray Bursts}\label{bursts}

We have analysed the six X-ray bursts of the 1999  
BeppoSAX observation. 
We do not have enough statistics to select different intervals during the 
bursts and analyse the rise and decay spectra separately. Therefore we have
analysed six spectra (one for each burst) each averaged over the whole $\sim
40$ s burst duration. 
From these spectra we have subtracted the spectrum of the persistent emission
and fitted them with a blackbody component. 
In Table~\ref{tabspecB} we show the results of the fits for each of the six
bursts. Note that, because the spectra of the bursts soften significantly
during the observation (see Fig.~1) the fit parameters, such as 
blackbody color temperatures, have to be considered average estimates over the
bursts.

\section{Discussion}

We have analysed two BeppoSAX observations (taken in 1999 and 2001) and RXTE
observations (taken in 1997 and 2000) of \src and discovered this source to be
an X-ray burster. This allows us to classify \src as a NS
system. In the 1999 BeppoSAX observation we
found six bursts. Five bursts have a rapid rise ($\la 5$~s) with a slow decay
($\sim 35$~s), and one has a
triangular shape, although it shows spectral properties similar to those of
the other bursts.
The fluxes of the persistent emission
in the RXTE observations and in the 2001 BeppoSAX observation are
$\sim7$ and $\sim 8$ times, respectively, higher than the flux of the
persistent emission in the 1999 BeppoSAX observation.
While we observe six bursts during the 1999 BeppoSAX
observation we do not observe any burst during the RXTE observations and
during the 2001 BeppoSAX observation. 
This would be in agreement with the general behaviour of X-ray bursters: 
the bursts disappear above a certain flux value due to steady nuclear burning
(e.g., van Paradijs et al. 1979; van Paradijs et al. 1988; see Lewin et
al. 1995 for a review).     

\begin{figure}
\centering
\begin{tabular}{c}
\psfig{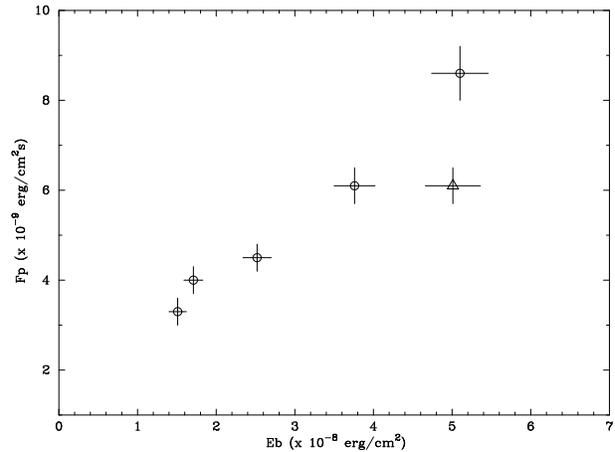}
\end{tabular} 
  \caption{Flux at the peak ($F_{p}$) as a function of the fluence ($E_{b}$)
of the six bursts. The empty triangle represents the burst with a triangular
shape.}   
  \label{F-E}
\end{figure}

Several sources show a correlation between the peak
flux and the fluence of the bursts (e.g. Sztajno et al. 1983; Basinska et
al. 1984; Lewin et al. 1987).   
Some of them also show a saturation in the peak flux at high fluences,
which has been explained by the fact that the luminosity has reached a
critical value that can be interpreted as the Eddington limit luminosity.
One of the best examples is MXB 1728-34 (Basinska et al. 1984; Di Salvo et
al. 2000; Galloway et al. 2002). This property can be used to infer the
distance $D$ to the source.   
In Fig.~\ref{F-E} we plot the flux at the peak $F_{p}$ (calculated using
the count rate at the peak of the burst in the 0.3~s resolution  
light curve) versus the fluence $E_{b}$ (the average bolometric flux 
times the duration of the burst) for each of the six bursts of \sr. The 
plot shows an approximately linear correlation between $F_{p}$ and $E_{b}$. 
Although we do not see a saturation of $F_{p}$ in this correlation
(and therefore we cannot derive the Eddington luminosity and the distance to
the source) we can at least give an upper limit to the distance assuming the
highest peak flux of the bursts to be less than $L_{Edd}$.  
Since $F_{p}\times 4\pi D^{2}\la L_{Edd}\sim
2.5\times10^{38}$ erg s$^{-1}$ (assuming ${\rm M}=1.4~\Msun$ and correcting
for gravitational redshift; van Paradijs \& McClintock 1994), taking
the bolometric flux at the peak of the first burst ($F_{p}=8.6\times 10^{-9}$
erg~cm$^{-2}$s$^{-1}$),
we obtain ${\rm D}\la 16$~kpc, not a stringent constraint.
Since the source is in the direction of the galactic center we assume
8~kpc as the source distance. At 8~kpc saturation would occur at 4 times this
flux level.    
 
\begin{figure}
\centering
\begin{tabular}{c}
\psfig{figure=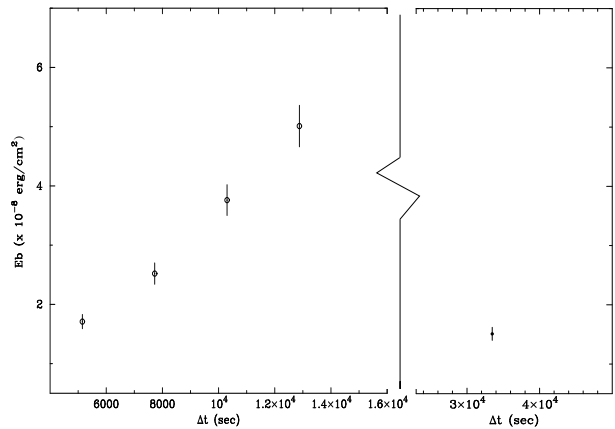,width=8.0cm,angle=0.0}
\end{tabular} 
 \caption{Fluence E$_{b}$ of
five bursts (second, third, fourth, fifth: open circles; sixth: filled
circle) as a function of the time interval from the previous burst.}
 \label{F-nP}
\end{figure}

Note that, as was pointed out by many authors, sometimes not all the accreted
fuel is burned during the burst event, and  the presence of a residual fuel
can have significant implications for the lack of regularity of the burst
behaviour (Ayasly \& Joss 1982; Hanawa \& Fujimoto 1984; Woosley \& Weaver
1985; Lewin et al. 1995).   
To investigate this point we have also  calculated the values of
$\alpha$=(GM/R)/$E_{nuc}$ 
(an observational quantity defined as the ratio of the average total 
energy in the persistent emission to that emitted in the burst) which is
expected to be $\sim 40$ for a thermonuclear burst in the mixed
hydrogen/helium burning regime ({\it e.g.}, Lewin et al. 1995). We measure
$\alpha \sim 40$ for all the bursts, but one, starting from the first
burst. Given that the accretion rate is consistent with being constant,
this implies a correlation between the time interval during which
matter accretes onto the NS and the energy emitted in the following
burst. The only 
exception is the last burst for which, assuming that the fifth and the sixth
bursts are  consecutives, we find a value for $\alpha$ that is much larger
than 40 ($\alpha \sim 250$).
In Fig.~\ref{F-nP} we plot the fluence 
of the bursts as a function of the time interval between each burst and the
previous one. There is a clear trend of the fluence to increase with the
time interval for all the bursts, except the sixth. This trend is in agreement 
with the behaviour of other sources (Lewin et al. 1995 for a review) and with
the expectation that the longer the burst interval is, the larger is the
amount of nuclear fuel available for the burst. The sixth burst seems not to
follow this trend. Most probably we have missed some bursts in between the
fifth and the sixth because of the presence of gaps (see \S 2). Indeed,
we find a gap about 3000~s before the sixth burst, just where, following
the correlation in Fig.~\ref{F-nP}, we would expect to find another burst.   

Based on the observed persistent flux and the distance to the source, we can
estimate the accretion rate and therefore the burning regime expected in this
case.
We find $\dot{\rm M} \sim 7\times 10^{-11}~\Msun$/yr. This corresponds
to the mixed hydrogen/helium burning regime triggered by thermally unstable
hydrogen (Fujimoto et al. 1981; Fushiki and Lamb 1987; see also Bildsten
2000). This is consistent with the value of $\alpha\sim 40$ we find for almost
all the bursts.

\section*{Acknowledgements}
SM acknowledges financial support from the Nederlandse Organisatie voor
Wetenschappelijk Onderzoek. TB acknowledges the CARIPLO foundation for
financial support. WHGL is grateful for support from NASA.
We would like to thank the anonymous referee for helpful comments on the
paper.

\end{document}